\documentclass[12pt]{iopart}
\usepackage{iopams}
\usepackage{latexsym}

\newcommand{\ii}{\mathrm{i}}
\newcommand{\dd}{\mathrm{d}}
\newcommand{\eqref}[1]{(\ref{#1})}

\begin{document}

\title{Quantum quenches in integrable systems: Constraints from factorisation}

\author{Dirk Schuricht}

\address{Institute for Theoretical Physics, Center for Extreme Matter and Emergent Phenomena, Utrecht University, Leuvenlaan 4, 3584 CE Utrecht, The Netherlands}

\ead{d.schuricht@uu.nl}

\begin{abstract}
We consider quantum quenches in integrable systems where complete factorisation of scattering, transmission and particle creation processes is assumed at all times. We show that under this assumption, the simultaneous transmission and creation of particles is impossible in generic interacting theories.
\end{abstract}

\maketitle

\section{Introduction}\label{sec:intro}
The unitary time evolution of closed quantum systems is one of the oldest problems in quantum mechanics. In its modern appearance as a sudden quantum quench~\cite{CalabreseCardy06} a system is prepared in an eigenstate, often the ground state, of the pre-quench Hamiltonian. At time $t=0$ one or more of the system parameters are changed suddenly, thus causing a non-trivial time evolution governed by the post-quench Hamiltonian. In the past years the tremendous progress in the creation and manipulation of ultra-cold systems~\cite{Bloch-08} has brought this setting to the laboratory. From the theoretical point of view the problem is straightforward to state: determine the initial state, calculate its time evolution under the post-quench Hamiltonian and evaluate the observables of interest. However, actually performing these steps is much harder (see References~\cite{Polkovnikov-11} for an overview of the theoretical developments). One particular class of systems for which progress has been made (and on which we focus here) are integrable models, as the integrability offers theoretical tools to handle the steps mentioned above. The developed approaches include the direct numerical calculations based on the Bethe-ansatz solutions~\cite{Faribault-09JSM}, the representation of the time-evolved wave function via contour integrals~\cite{IyerAndrei12}, thermodynamic arguments employed in the quench-action method~\cite{CauxEssler13,BSE14} or the application of form-factor expansions~\cite{Gritsev-07prl,FiorettoMussardo10,KormosPozsgay10,BSE14}. 

The latter approach relies on the knowledge of the scattering theory of the excitations, which must be factorisable in the sense that multi-particle scattering processes can be decomposed into successive two-particle events. Thus in the context of quantum quenches the natural question arises which quench transformations will preserve the underlying structure of the scattering theory. This question was addressed by Sotiriadis et al~\cite{Sotiriadis-12} for systems with a single particle species. They showed that the well-known linear Bogoliubov transformations are restricted to free theories and identified two classes of possible transformations for interacting theories. 

Recently, Delfino~\cite{Delfino14} investigated the constraints following from factorisability, again considering systems with a single particle species, and showed that the only consistent theories are provided by the free boson and free fermion. One may wonder whether the existence of many particle species makes non-trivial, factorisable quenches possible. The aim of the present article is to address this question. We will first define the concept of factorisability at all times and present the constraints originating from it. We then analyse the consequences of these constraints and show that in generic interacting theories [satisfying \eqref{eq:Sint}] the simultaneous existence of particle transmission and pair creation is impossible. Finally we consider the sine-Gordon model as an explicit example.

\section{Factorised quenches}\label{sec:factorisedquenches}
\begin{figure}[t]
\begin{center}
\setlength{\unitlength}{10pt}
\begin{picture}(42,14)(0,0)
\put(0,13){(a)}
\put(2,2){\line(1,1){8}}\put(1,10.7){$\theta_2, b_2$}\put(9.5,0.5){$\theta_2, a_2$}
\put(10,2){\line(-1,1){8}}\put(9.5,10.7){$\theta_1, b_1$}\put(1,0.5){$\theta_1, a_1$}
\put(7,5.8){$S_{a_1a_2}^{b_1b_2}(\theta_1-\theta_2)$}

\put(17,13){(b)}
\multiput(19,6)(1,0){8}{\line(1,0){0.5}}
\put(22,2){\line(1,4){2}}\put(21.5,0.5){$\xi, a$}\put(24,10.7){$\theta,b$}
\put(23.2,4.5){$T_a^b(\theta)$}
\put(19,6.3){$t=0$}

\put(30,13){(c)}
\multiput(32,6)(1,0){10}{\line(1,0){0.5}}
\put(37,6){\line(-1,1){4}}\put(31,10.7){$-\theta,b_1$}
\put(37,6){\line(1,1){4}}\put(40,10.7){$\theta,b_2$}
\put(35.5,4.5){$K^{b_1b_2}(\theta)$}
\put(32,6.3){$t=0$}
\end{picture}
\end{center}
\caption{(a) Graphical depiction of a scattering process of two incoming particles of species $a_{1,2}$ with rapidities $\theta_{1,2}$ into outgoing particles of species $b_{1,2}$. (b) Transmission of a particle parametrised via the post-quench rapidity $\theta$. The dashed line indicates the quench at $t=0$. (c) Creation of a pair of particles at $t=0$.}
\label{fig:definitions}
\end{figure}
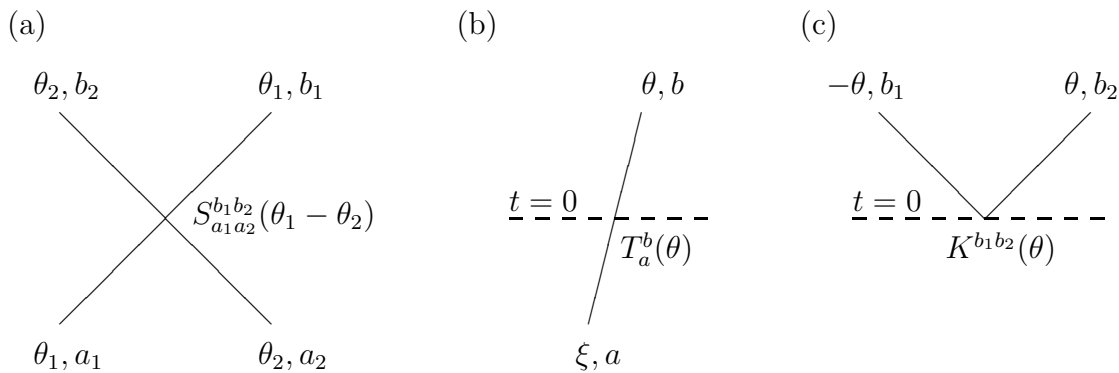

We consider a (1+1)-dimensional field theory with particles created and annihilated by the Faddeev--Zamolodchikov operators~\cite{ZamolodchikovZamolodchikov79,Faddeev80} $Z_a^\dagger(\theta)$ and $Z_a(\theta)$, where $a$ labels the particle species. The rapidity $\theta$ encodes its energy and momentum via $E=\Delta\cosh\theta$ and $p=\frac{\Delta}{v}\sinh\theta$ with the particle mass $\Delta$ and velocity $v$ respectively. The scattering theory of the particles is assumed to be completely elastic and factorisable, ie, any many-particle scattering process can be factorised into successive two-particle processes described by the scattering matrix $S_{a_1a_2}^{b_1b_2}(\theta_1-\theta_2)$ depicted in figure~\ref{fig:definitions}(a). In this sense the theory is integrable~\cite{Mussardo10,CauxMossel11}. Furthermore we assume the theory to respect C, P and T, ie, $S_{a_1a_2}^{b_1b_2}(\theta)=S^{a_1a_2}_{b_1b_2}(\theta)=S_{a_2a_1}^{b_2b_1}(\theta)=S_{\bar{a}_1\bar{a}_2}^{\bar{b}_1\bar{b}_2}(\theta)$, where the bar denotes the charge conjugated index.

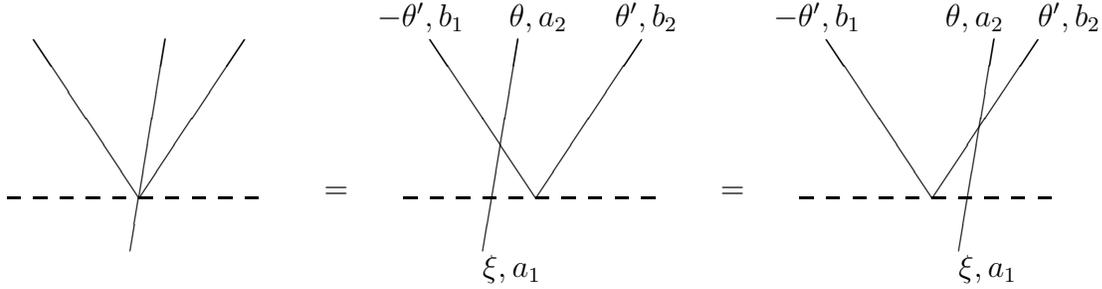
\begin{figure}[t]
\begin{center}
\setlength{\unitlength}{10pt}
\begin{picture}(42,12)(0,0)
\multiput(0,3)(1,0){10}{\line(1,0){0.5}}
\put(5,3){\line(-2,3){4}}
\put(5,3){\line(2,3){4}}
\put(4.65,1){\line(1,6){1.34}}

\put(12,3){=}
\multiput(15,3)(1,0){10}{\line(1,0){0.5}}
\put(20,3){\line(-2,3){4}}\put(14,9.5){$-\theta', b_1$}
\put(20,3){\line(2,3){4}}\put(23,9.5){$\theta', b_2$}
\put(18,1){\line(1,6){1.34}}\put(18,0){$\xi,a_1$}\put(19,9.5){$\theta,a_2$}

\put(27,3){=}
\multiput(30,3)(1,0){10}{\line(1,0){0.5}}
\put(35,3){\line(-2,3){4}}\put(29,9.5){$-\theta', b_1$}
\put(35,3){\line(2,3){4}}\put(39,9.5){$\theta', b_2$}
\put(36,1){\line(1,6){1.34}}\put(36,0){$\xi,a_1$}\put(35.5,9.5){$\theta,a_2$}
\end{picture}
\end{center}
\caption{The factorisability at $t=0$ implies that the four-particle process on the left-hand side can be rewritten as one transmission and one pair creation. The second equality yields the consistency relation \eqref{eq:relation1}.}
\label{fig:relation1}
\end{figure}

In such a theory we consider a quantum quench~\cite{CalabreseCardy06} at time $t=0$, ie, the sudden, global change of some parameter (we assume translational invariance of the theory at all times). All processes in the system are assumed to be factorisable before and after the quench but also at $t=0$, a situation we call a ``factorised quench". Translational invariance and factorisation imply that particles present at $t<0$ will preserve their momentum through the quench. We denote the corresponding transmission amplitude by $T_a^b(\theta)$ and parametrise it using the post-quench rapidity $\theta$, see figure~\ref{fig:definitions}(b). If the particle mass and velocity take the pre-quench values $\Delta'$ and $v'$ respectively, the pre-quench rapidity $\xi$ is fixed by $\frac{\Delta}{v}\sinh\theta=\frac{\Delta'}{v'}\sinh\xi$. In addition, the quench may create (or annihilate) sets of particles with total momentum zero. The simplest process is the creation of a pair of particles with rapidities $\theta$ and $-\theta$, the corresponding creation amplitude $K^{b_1b_2}(\theta)$ is depicted in figure~\ref{fig:definitions}(c). We stress that the assumption of factorisability implies that all processes at $t=0$ can be decomposed into individual transmission and creation (or annihilation) amplitudes, see figure~\ref{fig:relation1} for an example. In particular, it is always possible to disentangle the creation and transmission of particles.

In the following section we derive several consistency relations for the scattering, transmission and creation of particles following from the assumption of factorisation at all times. In the case of a single particle species this problem has been analysed by Delfino~\cite{Delfino14}. 

\section{Consistency relations}\label{sec:relations}
\begin{figure}[b]
\begin{center}
\setlength{\unitlength}{10pt}
\begin{picture}(35,12)(0,0)
\multiput(5,3)(1,0){10}{\line(1,0){0.5}}
\put(9,3){\line(-1,3){2}}\put(4,9.5){$-\theta', b_1$}
\put(9,3){\line(1,3){2}}\put(9.5,9.5){$\theta', b_2$}
\put(5.5,1){\line(1,1){8}}\put(4,0){$\xi,a_1$}\put(13.3,9.5){$\theta,a_2$}
\put(18,3){=}
\multiput(22,3)(1,0){10}{\line(1,0){0.5}}
\put(25,3){\line(-1,3){2}}\put(20,9.5){$-\theta', b_1$}
\put(25,3){\line(1,3){2}}\put(27,9.5){$\theta', b_2$}
\put(24.5,1){\line(1,1){8}}\put(23,0){$\xi,a_1$}\put(31.5,9.5){$\theta,a_2$}
\end{picture}
\end{center}
\caption{Graphical representation of the relation \eqref{eq:relation2}.}
\label{fig:relation2}
\end{figure}
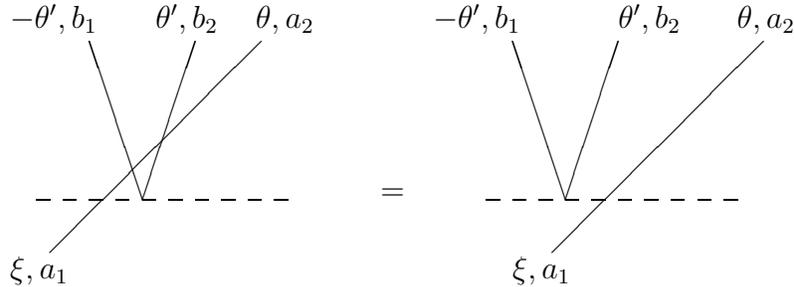
To obtain the first consistency relation we consider a process in which a single particle with rapidity $\theta$ is transmitted while an additional pair of particles with rapidities $\pm\theta'$ is created, see figure~\ref{fig:relation1}. The assumption of factorisability implies that the total amplitude can be expressed in terms of $T_{a_1}^{a_2}(\theta)$ and $K^{b_1b_2}(\theta')$ alone, and that it is not affected by translations of the particle trajectories. Hence, if we assume $0<\theta<\theta'$ and translate the trajectory of the transmitted particle as depicted, we obtain the requirement 
\begin{equation}
S_{c_1c_2}^{a_2b_1}(\theta+\theta')\,T_{a_1}^{c_1}(\theta)\,K^{c_2b_2}(\theta')=
S_{c_2c_1}^{b_2a_2}(\theta'-\theta)\,K^{b_1c_2}(\theta')\,T_{a_1}^{c_1}(\theta).
\label{eq:relation1}
\end{equation}
In the case of a single particle species the transmission and pair creation amplitudes drop out and we find the scattering to be independent of momentum~\cite{Delfino14}. Unitarity then implies $S^2=1$, ie, the only possibilities are the free boson and free fermion, for which the explicit details can be worked out via a Bogoliubov transformation (see, eg,~\cite{CalabreseCardy07,Sotiriadis-12}). The second consistency relation is obtained by considering $0<\theta'<\theta$ as shown in figure~\ref{fig:relation2}, it reads
\begin{equation}
S_{c_4c_2}^{a_2b_2}(\theta-\theta')\,S_{c_3c_1}^{c_4b_1}(\theta+\theta')\,T_{a_1}^{c_3}(\theta)\,K^{c_1c_2}(\theta')=
K^{b_1b_2}(\theta')\,T_{a_1}^{a_2}(\theta).
\label{eq:relation2}
\end{equation}
Multiplying by $S_{a_2b_2}^{d_1d_2}(\theta'-\theta)$ and using the unitarity of the scattering matrix this can be easily shown to be equivalent to relation \eqref{eq:relation1}. 

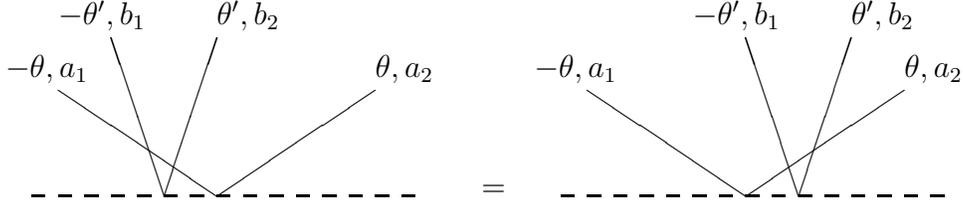
\begin{figure}[t]
\begin{center}
\setlength{\unitlength}{10pt}
\begin{picture}(35,12)(0,0)
\multiput(1,3)(1,0){15}{\line(1,0){0.5}}
\put(6,3){\line(-1,3){2}}\put(2,9.5){$-\theta', b_1$}
\put(6,3){\line(1,3){2}}\put(8,9.5){$\theta', b_2$}
\put(8,3){\line(-3,2){6}}\put(0,7.5){$-\theta, a_1$}
\put(8,3){\line(3,2){6}}\put(14,7.5){$\theta, a_2$}

\put(18,3){=}
\multiput(21,3)(1,0){15}{\line(1,0){0.5}}
\put(30,3){\line(-1,3){2}}\put(26,9.5){$-\theta', b_1$}
\put(30,3){\line(1,3){2}}\put(32,9.5){$\theta', b_2$}
\put(28,3){\line(-3,2){6}}\put(20,7.5){$-\theta, a_1$}
\put(28,3){\line(3,2){6}}\put(34,7.5){$\theta, a_2$}
\end{picture}
\end{center}
\caption{Factorisation condition \eqref{eq:relation3} involving two pair creation amplitudes, which is equivalent to the boundary Yang--Baxter equation in factorisable boundary field theories~\cite{GhoshalZamolodchikov94}.}
\label{fig:relation3}
\end{figure}
\begin{figure}[b]
\begin{center}
\setlength{\unitlength}{10pt}
\begin{picture}(30,12)(0,0)
\multiput(0,3)(1,0){8}{\line(1,0){0.5}}
\put(4,7){\line(-1,1){2}}\put(0,9.5){$-\theta, a_1$}
\put(4,7){\line(1,1){2}}\put(5.5,9.5){$\theta, a_2$}
\qbezier(4,3)(2,5)(4,7)\qbezier(4,3)(6,5)(4,7)

\put(10,3){=}
\multiput(14,3)(1,0){15}{\line(1,0){0.5}}
\put(21,3){\line(-1,1){6}}\put(13,9.5){$-\theta, a_1$}
\put(21,3){\line(1,1){6}}\put(26,9.5){$\theta, a_2$}
\end{picture}
\end{center}
\caption{Depiction of condition \eqref{eq:relation4}, which is equivalent to the boundary cross-unitarity condition in factorisable boundary field theories~\cite{GhoshalZamolodchikov94}.}
\label{fig:relation4}
\end{figure}
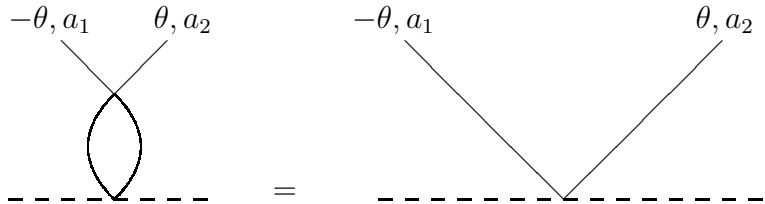
Furthermore, there are two relations involving the pair-creation amplitude only. The first is depicted in figure~\ref{fig:relation3}; explicitly it is given by
\begin{eqnarray}
&&S_{c_1c_4}^{b_1a_1}(\theta-\theta')\,S_{c_2c_3}^{b_2c_4}(\theta+\theta')\,K^{c_1c_2}(\theta')\,K^{c_3a_2}(\theta)\nonumber\\[2mm]
&&\qquad=S_{c_4c_3}^{a_2b_2}(\theta-\theta')\,S_{c_1c_2}^{c_4b_1}(\theta+\theta')\,K^{a_1c_1}(\theta)\,K^{c_2c_3}(\theta'),
\label{eq:relation3}
\end{eqnarray}
while the second relation reads (see figure~\ref{fig:relation4})
\begin{equation}
S^{a_1a_2}_{b_1b_2}(2\theta)\,K^{b_2b_1}(-\theta)=K^{a_1a_2}(\theta).
\label{eq:relation4}
\end{equation}
These relations also appear when studying the quench dynamics in integrable systems, where they emerge from the requirement that the initial state is well-defined and compatible with integrability~\cite{FiorettoMussardo10,BSE14}. In this context the quench setup is mapped to a corresponding boundary field theory in which  \eqref{eq:relation3} becomes the boundary Yang--Baxter equation while \eqref{eq:relation4} is the boundary cross-unitarity condition~\cite{GhoshalZamolodchikov94}. 

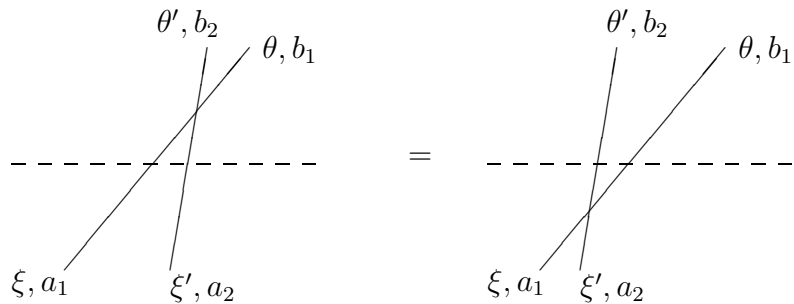
\begin{figure}[t]
\begin{center}
\setlength{\unitlength}{10pt}
\begin{picture}(30,12)(0,0)
\multiput(0,5)(1,0){12}{\line(1,0){0.5}}
\put(2,1){\line(5,6){7}}\put(6,0){$\xi',a_2$}\put(5.5,10){$\theta',b_2$}
\put(6,1){\line(1,6){1.4}}\put(0,0.2){$\xi,a_1$}\put(9.5,9){$\theta,b_1$}

\put(15,5){=}
\multiput(18,5)(1,0){12}{\line(1,0){0.5}}
\put(20,1){\line(5,6){7}}\put(18,0.2){$\xi,a_1$}\put(27.5,9){$\theta,b_1$}
\put(21.5,1){\line(1,6){1.4}}\put(21.5,0){$\xi',a_2$}\put(22.5,10){$\theta',b_2$}
\end{picture}
\end{center}
\caption{Consistency condition \eqref{eq:relation5} relating the scattering matrices before and after the quench.}
\label{fig:relation5}
\end{figure}

Finally, the assumption of factorised scattering before the quench yields a relation involving the transmission amplitude only, namely
\begin{equation}
S_{c_1c_2}^{b_1b_2}(\theta-\theta')\Big|_{t>0}\,T_{a_1}^{c_1}(\theta)\,T_{a_2}^{c_2}(\theta')=T_{c_2}^{b_2}(\theta')\,T_{c_1}^{b_1}(\theta)\,S_{a_1a_2}^{c_1c_2}(\xi-\xi')\Big|_{t<0}.
\label{eq:relation5}
\end{equation}
This relates the scattering matrices before and after the quench (see figure~\ref{fig:relation5}). In particular, note that the scattering matrix on the right-hand side depends on the pre-quench rapidities $\xi$ and $\xi'$. We have stated \eqref{eq:relation5} here for completeness, but the analysis in the following sections will be based solely on the relations \eqref{eq:relation1}--\eqref{eq:relation4}.

Finally we note that in the presence of bound states additional relations follow from the bootstrap principle~\cite{Mussardo10}. We also note that relations between scattering, transmission and reflection amplitudes similar to \eqref{eq:relation1}--\eqref{eq:relation5} have been obtained from the requirement of factorisability in integrable impurity systems with reflection and transmission~\cite{Delfino-94}.

\section{Diagonal theories}\label{sec:diagonal}
As already mentioned, in theories possessing only a single particle species the consistency relations given above are only compatible with free theories~\cite{Delfino14}. The next simplest situation we can think of is diagonal scattering, ie, 
\begin{equation}
S_{a_1a_2}^{b_1b_2}(\theta)=f_{a_1a_2}(\theta)\,\delta_{a_1}^{b_1}\,\delta_{a_2}^{b_2}.
\end{equation}
Inserting this into \eqref{eq:relation1} and \eqref{eq:relation2} implies $f_{a_1a_2}(\theta)=1$ or $f_{a_1a_2}(\theta)=-1$ independently of momentum and particles species. Therefore, we again find non-interacting theories. 

\section{General solution for non-diagonal theories}\label{sec:general}
In this section we show that provided the scattering matrix satisfies
\begin{equation}
S_{a_1a_2}^{b_1b_2}(\theta=0)=-\delta_{a_1}^{b_2}\,\delta_{a_2}^{b_1},
\label{eq:Sint}
\end{equation}
the consistency relations \eqref{eq:relation1}--\eqref{eq:relation4} cannot be satisfied simultaneously. Physically the opposite pairing of particle species and momentum in \eqref{eq:Sint} means that the excitations behave like impenetrable particles in collisions at small relative momentum. This is for example the case in the repulsive sine-Gordon model (see next section) or the O($n$) non-linear sigma model~\cite{ZamolodchikovZamolodchikov79,Mussardo10}. The behaviour \eqref{eq:Sint} is an essential ingredient for semi-classical treatments~\cite{SachdevDamle97} and as such has been used as a starting point in the study of quantum quenches in interacting systems~\cite{Evangelisti13}.

In order to make progress we start with relation \eqref{eq:relation2}. We assume the pre- and post-quench theories to possess the same set of particles species. Because of momentum conservation the transmission matrix is invertible at finite rapidity, ie, we can eliminate it yielding 
\begin{equation}
S_{c_3c_2}^{a_2b_2}(\theta-\theta')\,S_{a_1c_1}^{c_3b_1}(\theta+\theta')\,K^{c_1c_2}(\theta')=\delta_{a_1}^{a_2}\,K^{b_1b_2}(\theta').
\end{equation}
Now setting $\theta=-\theta'$ and using \eqref{eq:Sint} as well as \eqref{eq:relation4} we obtain
\begin{equation}
\delta_{a_1}^{a_2}\,K^{b_1b_2}(\theta')+\delta_{a_1}^{b_1}\,K^{b_2a_2}(-\theta')=0.
\end{equation}
This immediately implies $K^{b_1b_2}(\theta')=0$, ie, for factorised quenches with scattering matrix satisfying \eqref{eq:Sint} there is no consistent solution of \eqref{eq:relation1}--\eqref{eq:relation4} with simultaneous pair creation and transmission amplitudes. A similar conclusion was reached for integrable impurity systems with reflection and transmission~\cite{Delfino-94}.

We note that the sole existence of pair creation governed by the relations \eqref{eq:relation3} and \eqref{eq:relation4} is not excluded, as these relations are satisfied in any integrable field theory with a boundary~\cite{GhoshalZamolodchikov94}. For example, explicit results for the boundary reflection matrix and thus the pair creation amplitude have been derived for the sine-Gordon model~\cite{GhoshalZamolodchikov94,Caux-03} and the non-linear sigma model~\cite{Ghoshal94-2}. 

Furthermore,  we recall that the existence of non-factorisable processes at $t=0$ will render the above analysis incomplete, since more complicated processes like the dressing of transmitted particles may occur~\cite{Sotiriadis-12}. 

\section{Example: Sine-Gordon model}\label{sec:SGM}
To give an explicit example, let us consider the quantum sine-Gordon model 
\begin{equation}
H_{\rm SG}=\frac{v}{16\pi}\int dx\left[\big(\partial_x\Phi\big)^2+
\frac{1}{v^2}\big(\partial_t\Phi\big)^2\right]-\lambda\int dx\,\cos(\beta\Phi)
\label{eq:SGM}
\end{equation}
in the interacting, repulsive regime $1/2<\beta^2<1$. In this regime the elementary excitations are massive, relativistic solitons and antisolitons for which the index of the Faddeev--Zamolodchikov operators takes the values $a=\pm$. The non-vanishing elements of the scattering matrix are given by~\cite{ZamolodchikovZamolodchikov79,Mussardo10}
\begin{eqnarray}
S^{++}_{++}(\theta)&=&S^{--}_{--}(\theta)=S_0(\theta)=
-\exp\left[\ii\int\limits_0^\infty\frac{\dd t}{t}
\sin\!\left(\frac{t\theta}{\pi\xi}\right)\frac{\sinh\big(\frac{\xi-1}{2\xi}t\big)}
{\sinh\bigl(\frac{t}{2}\bigr)\,\cosh\big(\frac{t}{2\xi}\big)}\right],\nonumber\\
S^{+-}_{+-}(\theta)&=&S^{-+}_{-+}(\theta)
= -\frac{\sinh\big(\frac{\theta}{\xi}\big)}
{\sinh\big(\frac{\theta-\ii\pi}{\xi}\big)}\,S_0(\theta),\nonumber\\
S^{+-}_{-+}(\theta)&=&S^{-+}_{+-}(\theta)= -\frac{\ii\sin\big(\frac{\pi}{\xi}\big)}
{\sinh\big(\frac{\theta-\ii\pi}{\xi}\big)}\,S_0(\theta),
\label{eq:SGSmatrix}
\end{eqnarray}
where we have defined $\xi=\beta^2/(1-\beta^2)>1$. Obviously the scattering matrix satisfies \eqref{eq:Sint}. The sine-Gordon model with a boundary is integrable~\cite{MacIntyre95}; the known~\cite{GhoshalZamolodchikov94,Caux-03} solution of the boundary Yang--Baxter equation yields the general solution of \eqref{eq:relation3}
\begin{eqnarray}
K^{\pm\mp}(\theta)&=&\cos\!\left(\vartheta\mp\frac{\pi}{2\xi}\mp\frac{\ii\theta}{\xi}\right)\,K_0(\theta),\nonumber\\
K^{\pm\pm}(\theta)&=&\frac{\eta}{2}\sin\!\left(\frac{\pi}{\xi}+\frac{2\ii\theta}{\xi}\right)\,K_0(\theta).
\label{eq:SGKmatrix}
\end{eqnarray}
Here $\eta$ and $\vartheta$ are free parameters and $K_0(\theta)$ is an arbitrary function. It is now straightforward to check that the condition \eqref{eq:relation1} yields $T_a^b(\theta)=0$ in agreement with the general finding of the previous section.

This implies that it is impossible to perform a quench in the sine-Gordon model \eqref{eq:SGM}, say in the parameter $\lambda$, such that scattering, pair creation and transmission are fully factorisable at all times. However, quenches that violate the factorisation at $t=0$, or initial states without solitons or antisolitons and thus no particle transmissions are possible. The time evolution for the latter situation was investigated in References~\cite{Gritsev-07prl,BSE14}, where the initial state was assumed to correspond to Dirichlet boundary conditions $\Phi(t=0,x)=0$, which can be thought of as arising from an initial system with $\lambda|_{t<0}=\infty$. 

\section{Conclusion}\label{sec:conclusion}
We have investigated factorised quantum quenches in integrable systems, ie, quenches for which the scattering, transmission and creation of particles is factorisable at all times. We showed that the simultaneous existence of transmission and pair creation processes described by the amplitudes $T_a^b(\theta)$ and $K^{b_1b_2}(\theta)$ respectively is only possible in non-interacting theories with trivial scattering matrix $S_{a_1a_2}^{b_1b_2}(\theta)=\pm\delta_{a_1}^{b_1}\delta_{a_2}^{b_2}$. Hence factorised quantum quenches in interacting theories either possess vanishing transmission or pair creation amplitudes. 

The crucial assumption underlying our analysis is the factorisation of all processes at the quench time $t=0$. Dropping it makes more complicated processes like the dressing of transmitted particles possible, resulting in non-linear relations between the pre- and post-quench scattering theories. Such transformations have been investigated previously for systems with a single particle species~\cite{Sotiriadis-12}. An extension of their analysis to the more general setup considered here would be desirable.

\section*{Acknowledgement}
I would like to thank Vladimir Gritsev, Andrew Mitchell and Gabor Tak\'acs for useful comments. This work is part of the D-ITP consortium, a program of the Netherlands Organisation for Scientific Research (NWO) that is funded by the Dutch Ministry of Education, Culture and Science (OCW). This work was supported by the Foundation for Fundamental Research on Matter (FOM), which is part of the Netherlands Organisation for Scientific Research (NWO), under 14PR3168.

\section*{References} 


\end{document}